\documentclass[reqno]{amsart}

\usepackage{amsfonts}
\usepackage{bezier}
\usepackage{euscript}
\usepackage{amsfonts,amssymb,amsmath,amsbsy,amsthm,amscd}
\usepackage{xcolor}
\usepackage{graphicx}
\usepackage{color}
\usepackage[T2A]{fontenc}
\usepackage[utf8]{inputenc}
\usepackage[english]{babel}   

\def\dfrac#1#2{\displaystyle{#1\over #2}}

\begin{document}

\title[On the solution of the Kolmogorov-Feller equation
 ]{
On the solution of the Kolmogorov-Feller equation arising in the model of biological evolution
}


\author{Olga S. Rozanova}

\address{ Mathematics and Mechanics Department, Lomonosov Moscow State University, Leninskie Gory,
Moscow, 119991, Russian Federation}
\email{rozanova@mech.math.msu.su}

\subjclass{Primary 60E05; Secondary 35Q84; 	82C31}

\keywords{probability density, gene expression,  Kolmogorov-Feller equation, fundamental solution, exact solution
}

\begin{abstract}
The Kolmogorov-Feller equation for the probability density of a Markov process on a half-axis, which arises in important problems of biology, is considered. This process consists of random jumps distributed according to Laplace's law and a deterministic return to zero. It is shown that the Green's function for such an equation can be found both in the form of a series and in explicit form for some ratios of the parameters. This allows one to explicitly find solutions to the Kolmogorov-Feller equation for many initial data.
\end{abstract}

\maketitle



\bigskip
{\bf 1. Introduction and problem statement.}
  The cells of all living organisms contain three main macromolecules: DNA, mRNA and proteins. Matrix ribonucleic acid (mRNA) contains information about the primary structure (amino acid sequence) of proteins and plays an important role in gene expression. mRNA is synthesized from DNA during transcription, after which, in turn, it is used during translation as a template for protein synthesis.
Gene expression, that is, the process of transferring information from mRNA to proteins, consisting of a series of biochemical reactions that occur randomly inside living cells, has been studied from an experimental and theoretical point of view for half a century. However, the simplest mathematical model of protein distribution in a cell population depending on the protein concentration inside a particular cell 
was introduced only in 2006 in \cite{Friedman2006}. It assumes a stochastic spasmodic nature of gene expression according to an exponential law, accompanied by continuous deterministic degradation (reversion to zero).
Namely, the protein is produced in jumps, in which an mRNA molecule is translated into several protein molecules before disintegrating. The lifetime of an mRNA is considered to be short compared to the lifetime of a protein molecule; protein production occurs in random exponentially distributed uncorrelated events.

The probability density $P(t,x)\ge 0$ of such a Markov process is described by the following integro-differential equation \cite{Friedman2006}
\begin{eqnarray}\label{FPF}
&&{\frac {\partial }{\partial t}}P \left( t,x \right)  = {\frac {\partial }{\partial x}}
 \left(\beta\,x \,P \left( t,x \right)\right)  +
{\lambda}\, \left(
k\,\int _{0 }^{x }\!P \left( t,z \right) {{\rm e}^{-k
 \left( x-z \right) }}{dz}-P \left( t,x \right)  \right),\\
 &&\quad 0\le z \le x,\,t\ge 0,\nonumber
 \end{eqnarray}
where $\lim\limits_{x\to 0} x P(t,x)=0$ and $\beta, \lambda, k$ are positive constants.
This is a generalization of the Fokker-Planck-Kolmogorov equation, which is sometimes called the Kolmogorov-Feller equation. In the biological interpretation, the variable $x$ corresponds to the concentration of the protein inside a particular cell, $\beta$ is the rate of protein degradation, $\lambda$ is the rate of DNA transcription in mRNA, $k$ is the ratio of the rate of mRNA degradation to the rate of mRNA translation in protein molecules. The constants $\alpha=\frac{\lambda}{\beta}$ and $k$ are the main parameters characterizing protein production.

There is a very large number of works in which the \cite{Friedman2006} model is generalized, for example, \cite{Huang2014}, \cite{Bokes2021} and the references contained there, but the study of solutions of the \eqref{FPF} equation
  is limited to the study of stationary solutions and the asymptotics of solutions for large $x$. The dynamics of the solution in time, as a rule, is studied only numerically. In this communication, we want to show that the Green's function ${\mathcal G}(t,x,y)$ of the Cauchy problem, that is, the solution of the equation \eqref{FPF} with initial conditions
  \begin{equation}\label{ID}
  P|_{t=0}=\delta(x-y), \quad x\ge 0, \, 0\le y\le x,
 \end{equation}
can be found analytically in the form of a series, and for some relations between the parameters and in the form of a finite sum. This allows us to find a solution to the Cauchy problem for any integrable on the semi-axis and bounded initial conditions
 \begin{equation*}
P|_{t=0}=\phi(x)\ge 0, \quad \int\limits_{{\mathbb R}_+} \phi \, dx=1,
 \end{equation*}
as
 \begin{equation}\label{conv}
P(t,x)=\int _{0 }^{\infty }\!{\mathcal G} \left( t,x,y \right) \phi(y) {dy},
 \end{equation}
which is an explicit formula for some types of initial data. To ensure the classical smoothness of the solution, it is necessary to require
$\phi\in C^1(\overline{\mathbb R}_+)$.

\bigskip

{\bf 2. Finding the Green's function}

1. Applying the Laplace transform $x\to w$ to the \eqref{FPF} equation and the initial data \eqref{ID}, we obtain the Cauchy problem for $\mathcal L\{P \}=\mathcal L\{P \} \left( t,w \right)$
\begin{eqnarray*}\label{FPL}
{\frac {\partial }{\partial t}}\mathcal L\{P \}   + \beta w  {\frac {\partial }{\partial w}}
 \,\mathcal L\{P \}  +
 \frac{\lambda k}{w+k}\mathcal L\{P \}=0, \quad \mathcal L\{P \}|_{t=0}=\Theta (y) e^{-w y},
\end{eqnarray*}
whose solution has the form
\begin{eqnarray*}\label{FPLsol}
\mathcal L\{P \}(t,w)=\mathcal L\{{\mathcal G}\}(t,w,y)=
\left(\frac {w e^{-\beta t}+k }{w+k}\right)^{\alpha}\,e^{-y w e^{-\beta t}}, \quad \alpha= \frac{\lambda}{\beta}.
\end{eqnarray*}
Denote $\bar x= x-y e^{-\beta t}\ge 0 $. Note that $\frac {w e^{-\beta t}+k }{w+k}=1+W,$ where
$W=\frac {w (e^{-\beta t}-1) }{w+k}$, $|W|<1$. Then, expanding $\left(\frac {w e^{-\beta t}+k }{w+k}\right)^{\alpha}$ into a convergent binomial series and applying the inverse Laplace transform, we obtain
\begin{eqnarray}\label{G1}
\mathcal G(t,x,y)=\sum\limits_{i=0}^{\infty}  C^i_\alpha (e^{-\beta t}-1)^i \mathcal L^{-1}
\left\{\left(\frac {w }{w+k}\right)^i\right\} (t, \bar x).
\end{eqnarray}
Using the properties of the Laplace transform, we find that
\begin{eqnarray}\label{G2}
&&\mathcal L^{-1}
\left\{\left(\frac {w }{w+k}\right)^i\right\} (t, \bar x)= \mathcal L^{-1}
\left\{\left(1-\frac {k }{w+k}\right)^i\right\} (t, \bar x)=\\&&\delta(\bar x)+\sum\limits_{s=1}^i (-1)^s C^s_i \Psi_s(\bar x),
\nonumber\\
&&\Psi_s(\bar x)=\frac{1}{(s-1)!} k^s \bar x^{s-1} e^{-k \bar x}, \quad s\in \mathbb N.\nonumber
\end{eqnarray}
Substituting \eqref{G2} into \eqref{G1} and noticing that
  $1+\sum\limits_{i=1}^{\infty} C^i_\alpha (e^{-\beta t}-1)^i = e^{-\alpha \beta t} $,
  we obtain a representation of the solution in the form of a series converging for each $\bar x\in \mathbb R_+$ as a sum of the singular
  component
$\mathcal G_{sing}=A(t)\delta(\bar x)$ and the regular component $\mathcal G_{reg}$:
\begin{eqnarray}\label{Gfull}
&&\mathcal G(t,x,y)=e^{-\alpha \beta t}\,\delta(\bar x) +  e^{-k \bar x}\,
\sum\limits_{i=1}^{\infty}  C^i_\alpha (e^{-\beta t}-1)^i
\sum\limits_{s=1}^i C^s_i \frac{(-1)^s}{(s-1)!} k^s \bar x^{s-1},\\&&\quad i, s\in \mathbb N,\quad s\le i.\nonumber 
\end{eqnarray}
We see that the Green's function contains a singular component $\mathcal G_{sing}$ for all $t>0$, but its amplitude $A(t)\to 0$ for $t\to \infty$.

2.  If $\alpha=n\in \mathbb N$, then the sum \eqref{Gfull} becomes finite.

3. The regular component $\mathcal G_{reg}$ tends at $t\to \infty$ to the probability density of the gamma distribution,
\begin{eqnarray*}\label{Gst}
&&\mathcal G_{st}(x)=  e^{-k  x}\,
\sum\limits_{i=1}^{\infty}  C^i_\alpha (-1)^i
\sum\limits_{s=1}^i C^s_i \frac{(-1)^s}{(s-1)!} k^s  x^{s-1}= \frac{k^\alpha x^{\alpha-1}\, e^{-k x}}{\Gamma(\alpha)},\\&& \quad i, s\in \mathbb N,\quad s\le i,\quad x\ge 0,
\end{eqnarray*}
where $\Gamma(\alpha)$ is Euler's gamma function.
The stationary solution of the equation \eqref{FPF} of the form $\mathcal G_{st}(x)$ was already obtained in \cite{Friedman2006}. Its maximum at $\alpha\le 1$ is at the origin, and at $\alpha> 1$ it is at the point $x=\frac{\alpha-1}{k}$. 

Note that all the transformations were done formally, but after the explicit form of the Green's function is obtained, we see from \eqref{conv} that, under the conditions imposed above on the initial data, $P(t,x)$ is absolutely integrable on the half-axis function (due to the presence of the factor $e^{-kx}$), so the Laplace transform is defined. The inverse Laplace transform is also defined since the image is an analytic function.

\bigskip

{\bf 3. Examples.}
For some fairly wide classes of initial data for $\alpha=n\in \mathbb N$, it is possible to represent the solution of the Cauchy problem in the form of an explicit formula.
This, for example, $\phi(x)=A_1 x^{a_1}e^{-b_1 x}$,  $\phi(x)=A_2 x^{a_2}e^{-b_2 x^2}$, where $A_1, A_2, a_1, a_2, b_1, b_2$ are positive constants, chosen so as to ensure that  integral over the semi-axis is equal to one,
as well as piecewise constant or piecewise polynomial initial data. Note that discontinuities in the initial conditions do not smooth out, as happens in the case of the heat equation, but continue to be present for all $t>0$, but their amplitude tends to zero for $t\to \infty$. This happens due to the hyperbolicity of the equation (see below). Therefore, to extend the class of initial data to piecewise-smooth functions, one has to use the generalized formulation of the solution of  equation \eqref{FPF}.

For large $n$, the formulas can be quite cumbersome, but they are easily found using computer algebra packages. These formulas provide a large stock of tests for numerical methods for solving integro-differential equations.

As examples illustrating the dynamics of density, we consider the cases $n=1$ and $n=2$, for which the corresponding Green's functions $\mathcal G_1$ and $\mathcal G_2$ are written rather short. Namely,
 \begin{eqnarray*}
   \mathcal G_1(t,\bar x) &=& (1-e^{-\beta t}) e^{-k \bar x}+ e^{-\beta t}\delta(\bar x) \\
   \mathcal G_2(t,\bar x) &=& \left(2 k e^{-\beta t} (1-e^{-\beta t}) + k^2 (1-e^{-\beta t})^2 \bar x \right)\,e^{-k \bar x}+ e^{-2\beta t}\delta(\bar x).
 \end{eqnarray*}
Their limit behavior is significantly different: $\mathcal G_{1st}$ has a maximum at zero, while
  $\mathcal G_{2st}$ has a maximum at the point $x=\frac{1}{k}>0$.

As the initial data in both cases, we choose the function $\phi(x)= x e^{-{x}}$. In this case,  integral \eqref{conv}
can be elementary calculated. Fig.1 shows plots of the solution at different times for $n=1$ (left) and $n=2$ (right).
We see that for $n=1$ the density maximum tends monotonically in time to the origin, while for $n=2$ the density maximum first also tends to the origin, but then the graph has a competing maximum, which eventually tends to maximum of the stationary solution, while the first maximum vanishes.

\begin{figure}[htb]
\begin{minipage}{0.4\columnwidth}
\includegraphics[scale=0.6]{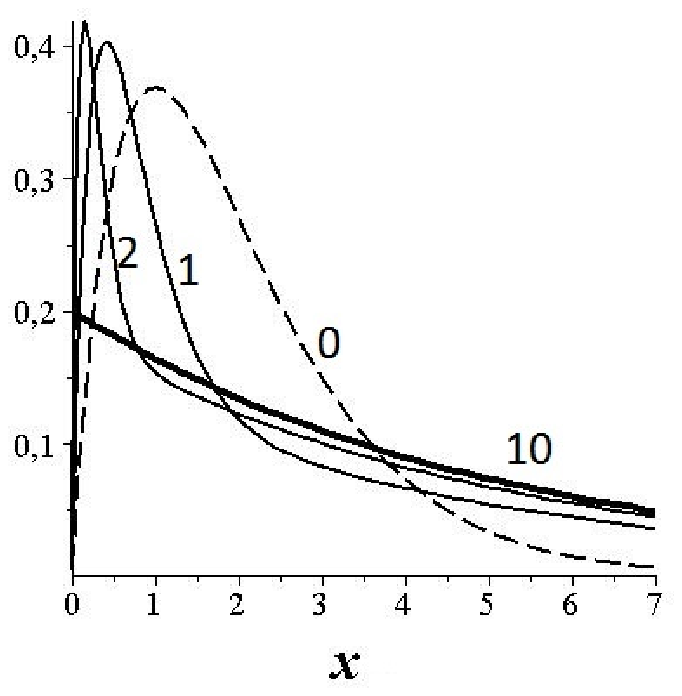}
\end{minipage}
\begin{minipage}{0.4\columnwidth}
\includegraphics[scale=0.6]{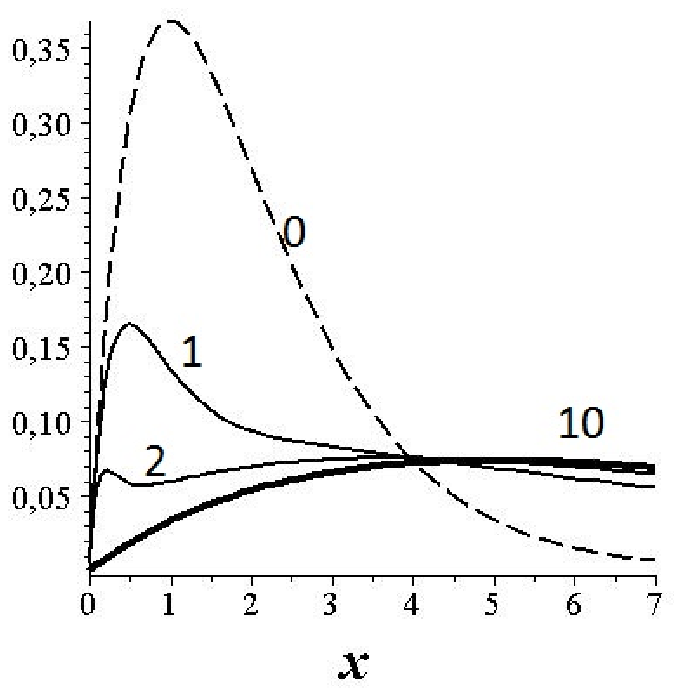}
\end{minipage}
\caption{Density for $n=1$ (left) and $n=2$ (right) for $t=0$ (dotted line), $t=1$, $t=2$, $t=10$ (thick line ); $k=0.2$, $\beta=1$.}\label{Pic1}
\end{figure}

\bigskip

 {\bf 4. Generalizations and remarks.} Equation \eqref{FPF} uses the simplest form of the density of jumps $p(z)=k e^{-kx}, x\ge 0$. Initially, it was chosen not only for reasons of simplicity, but also because of its compliance with experimental data. However, if we solve the purely mathematical problem of finding the Green's function, and hence the solutions of the Cauchy problem in explicit form, then we can consider wider classes of functions as the kernel $p(z)$. The solution can be obtained explicitly if $p(z)$ is a solution to a linear equation with constant coefficients of any order. For example, it can be a finite sum of exponents of the form $\frac{1}{j}\,\sum\limits_{i=1} ^j k_i e^{-k_i x},\, x\ge 0$, $j \in \mathbb N$.

Note that the method of finding the Green's function is standard, but the fact that it is possible to obtain an explicit expression with the inverse Laplace transform is a rather rare phenomenon.

As shown in \cite{Huang2014}, by replacing $Y(t,x)=\int\limits_0^x P(t,x-z) e^{-kz} dz$, the integro-differential equation \eqref{FPF} can be reduced to the differential equation
 \begin{eqnarray*}
\dfrac{\partial^2 Y}{\partial t \partial x} -\beta x \dfrac{\partial^2 Y}{\partial x^2} +k  \dfrac{\partial Y}{\partial t}+
(\lambda-\beta (kx+1))\dfrac{\partial Y}{\partial x}+k \beta Y=0, \quad x\ge 0,
 \end{eqnarray*}
belonging to the hyperbolic type. This explains the fact that discontinuities in the initial data do not disappear with time, but propagate along the characteristics. The characteristics are $x(t)=x_0 e^{-\beta t}$, $x_0\ge 0$, and $t=\rm const$. We see that one more family of characteristics is added to the "parabolic" one.


\end{document}